\documentclass{article}
\usepackage{dcase2023,amsmath,amssymb,graphicx,url,times,booktabs, tabularx, hyperref}

\title{Multi-label open-set audio classification}

\twoauthors
  {Sripathi Sridhar}
    {New Jersey Institute of Technology\\
     ss645@njit.edu}
  {Mark Cartwright}
    {  New Jersey Institute of Technology \\
     mark.cartwright@njit.edu}

\name{Sripathi Sridhar$\sthanks{This work is partially supported by NSF award \#1955357}$,
      Mark Cartwright}
\address{Sound Interaction and Computing (SInC) Lab, New Jersey Institute of Technology \\\{ss645, mark.cartwright\}@njit.edu}          

\begin{document}

\ninept
\maketitle

\begin{sloppy}

\begin{abstract}
Current audio classification models have small class vocabularies relative to the large number of sound event classes of interest in the real world. Thus, they provide a limited view of the world that may miss important yet unexpected or unknown sound events. To address this issue, open-set audio classification techniques have been developed to detect sound events from unknown classes. Although these methods have been applied to a multi-class context in audio, such as sound scene classification, they have yet to be investigated for polyphonic audio in which sound events overlap, requiring the use of multi-label models. In this study, we establish the problem of multi-label open-set audio classification by creating a dataset with varying unknown class distributions and evaluating baseline approaches built upon existing techniques.
\end{abstract}


\section{Introduction}
\label{sec:intro}

Audio classification (AC), the machine listening task of identifying sound events in an audio recording, has typically been studied as two task variants, i.e. multi-class AC, where the input recordings are expected to contain only one event, and multi-label AC, where the input recordings may contain multiple overlapping sound events. Real-world audio recordings in typical urban, domestic or environmental settings often contain multiple sound sources of anthrophony, biophony, and geophony, and thus, are better modeled as a multi-label AC task. 


Multi-label AC is a common machine listening task that has been applied to various scenarios such as urban sound data \cite{cartwright_sonyc_2019}, everyday environments \cite{cakir_polyphonic_2015}, and music \cite{gururani2018instrument}. Much of this work however assumes a small fixed class vocabulary, a closed-set task, which does not reflect real-world scenarios. Everyday sound scenes consist of sources drawn from hundreds if not thousands of classes depending on the class granularity of interest, and people are constantly exposed to novel classes, e.g., those from new or uncommon technology and animal vocalizations. To the ``ears'' of these models, unknown sound classes simply do not exist or --- possibly worse --- are confused with known classes. This limited class vocabulary size can be attributed to the cost and difficulty of annotating large-scale audio datasets. However, the result of this barrier is a limited view of the acoustic world by AC models that may miss important yet unexpected or unknown sound events, hindering machine listening's transformative potential.

One solution to this problem is to build models with a dynamic vocabulary that can be updated in a lightweight manner without having to retrain the model from scratch. An example of this approach is few-shot classification \cite{fink2004object}, which is often formulated within a meta-learning framework where a model can learn a new class from a small `support set' of examples \cite{snell2017prototypical}. Prior work has applied this to tasks such as instrument recognition \cite{flores2021leveraging}, multi-label audio classification \cite{wang_few-shot_2021}, and multi-label drum transcription \cite{wang_few-shot_2020}. However, this method still requires the user or researcher to supply a support set for unseen or novel classes \cite{wang_few-shot_2021}, and thus, such supervised approaches are only useful if you know what you are hoping to find and have examples of it. In many situations --- e.g., urban noise monitoring, audio accessibility, bioacoustic monitoring --- it is the rare events and unexpected events that are arguably the most important to detect, i.e., the machine listening equivalent of a ``black swan event'' \cite{taleb2007black}. To this end, we focus on detecting the presence of unknown classes in addition to known classes, referred to as open-set modeling.


Open-set modeling has seen research interest in the image domain for several years \cite{bendale_towards_2015, zhang2020hybrid}, but it has only more recently gained interest in the audio domain and been applied to tasks such as domestic sound classification \cite{naranjo-alcazar_open_2020}, acoustic scene classification \cite{kwiatkowska2020deep, mesaros2019acoustic}, and the related yet distinct task of anomalous sound detection  \cite{Kawaguchi_arXiv2021_01}. However, all of these tasks are binary or multi-class AC --- to the best of our knowledge, open-set modeling has not been applied to multi-label AC.

As in \cite{naranjo-alcazar_open_2020, bendale_towards_2015}, we define \textit{known known} (KK) classes as known (i.e., in-vocabularly) classes seen during training and inference, \textit{known unknown} (KU) classes as unknown (i.e., out-of-vocabulary) classes seen during training and inference, and \textit{unknown unknown} (UU) as unknown classes seen only at inference. A fourth category, unknown known (UK) classes, are classes in which only semantic or metadata information is available in the absence of discrete labels --- this category is not considered in this work. We collectively refer to KU and UU as unknown classes, and KK as known classes. 

We define multi-label open-set AC (MLOS) as the task of assigning between $0$ and $|\text{KK}|+1$ class labels to an audio recording, where $|\text{KK}|$ is the cardinality of the set of known classes and $+1$ refers to the label indicating the presence of an unknown sound class. Thus, an MLOS model needs to both estimate which known classes are present as well as decide whether at least one unknown class is present.  This is in contrast to multi-class open-set AC models which assign only 1 of $|\text{KK}|+1$ class labels to an audio recording.  

In this paper, we (1) establish the problem of MLOS, (2) introduce a new dataset with varying unknown class distributions to investigate this problem, and (3) evaluate baseline approaches comprised of combinations of existing machine listening techniques.

\section{Dataset}
\label{sec:data}

Prior open-set AC datasets are either multi-class \cite{naranjo-alcazar_open_2020} or focused on binary anomalous sound detection \cite{koizumi_unsupervised_2019}. In order to establish the MLOS task, we are interested in exploring the effects of polyphony, and levels of ``openness'' while working with a large class vocabulary. While few-shot datasets like FSD-MIX-CLIPS  \cite{wang_few-shot_2021} meet the polyphony criteria, they do not have varying levels of ``openness'' nor dataset variants where different classes are assigned to the KK, KU and UU categories. As in \cite{geng_recent_2020}, we define ``openness'' as 
\begin{equation}
    O^* = 1 - \sqrt{\left(2 \times C_{tr}\right)/ \left(C_{tr} + C_{te}\right)},
\end{equation}
 where $C_{tr} = |\text{KK} \cup \text{KU}|$ is the number of classes seen during training and $C_{te} = |\text{KK} \cup \text{KU} \cup \text{UU}|$ the number of classes seen during testing. 
Thus, for larger $C_{tr}$, we assign lower values of openness.

To this end, we develop a new dataset of synthetic soundscapes using open-set criteria. As in FSD-MIX-CLIPS, we use a subset of FSD50K where each clip has a single `present and predominant' label, i.e., the labeled sound event is the only type of sound present with the exception of mild background noise \cite{fonseca2022FSD50K}. This gives us 7600 source events from 89 classes, each between 0.5s and 4s in duration. 
We use only the leaf node labels according to the Audioset ontology \cite{gemmeke2017audio}. Hereafter we refer to this subset of FSD50K as the \textit{source dataset}. 

First, we split the classes into 5 subsets of 18 classes each (except for one subset with 17 classes), and from these subsets, we create 10 variations of class assignments into KK, KU, and UU as shown in Table\ref{tab:openness_table} --- 5 with a low degree of openness and 5 with a high degree of openness, i.e. no KU classes. The openness coefficients are $O^*=0.05$ or $0.06$ for low openness ($C_{tr}=72 \text{ or } 71$) and $O^*=0.13$ or $0.14$ for high openness ($C_{tr}=54 \text{ or } 53$).
For each class assignment variation $i$, we generate an intermediate dataset called `Open-Set Soundscape-i' (OSS-i), consisting of 10s 44.1kHz synthetic soundscapes using Scaper \cite{salamon_scaper_2017} --- 200k training, 30k validation, and 30k test with no source overlap between splits.  The training and validation sets are synthesized from only the known class subsets, e.g. in dataset variant 1, from L1-L4 in the low openness case and H1-H3 in the high openness case (Var. 1 in Table \ref{tab:openness_table}). In both openness cases, the test set is synthesized using all the subsets. Additionally, we also create a small tuning validation set using all the subsets for hyperparameter tuning, ensuring no example overlap with the test set.

In each OSS-i, we maintain the class distribution of the source dataset as closely as possible while enforcing a minimum of 200 examples per class. 
Each soundscape has one to four overlapping source sound events in the foreground, which we place between 0 to 9s in the soundscape. 
We augment each source with pitch shifting (-2 to +2 semitones) and time stretching (by a factor of 0.8 to 1.2).
We use uniform random sampling for all augmentations during generation. 
 
For each OSS-i dataset variant, we generate a dataset of 1s clips by centering a window on each event in the 10s soundscape and labeling a class as present if it overlaps with this window.  
This yields 10 datasets (5 high, 5 low openness) with $\sim$500k clips each. 

We refer to this as the Open-Set Tagging (OST) dataset and use it to train and evaluate our models. Both OSS and OST datasets are publicly available \footnote{\href{https://doi.org/10.5281/zenodo.7241704}{10.5281/zenodo.7241704}}.
\begin{table}[]
\setlength{\tabcolsep}{4pt}
{\footnotesize
\begin{tabular}{c|ccccc|ccccc}
Openness & \multicolumn{5}{c|}{Low} & \multicolumn{5}{c}{High} \\ \hline
Subset      & L1   & L2  & L3  & L4  & L5  & H1    & H2  & H3  & H4  & H5  \\ \hline
Var. 1 & KK  & KK & KK & KU & UU & KK   & KK & KK & UU & UU \\
Var. 2 & UU  & KK & KK & KK & KU & UU   & KK & KK & KK & UU \\
Var. 3 & KU  & UU & KK & KK & KK & UU   & UU & KK & KK & KK \\
Var. 4 & KK  & KU & UU & KK & KK & KK   & UU & UU & KK & KK \\
Var. 5 & KK  & KK & KU & UU & KK & KK   & KK & UU & UU & KK
\end{tabular}
}
\caption{Class splits for high and low openness dataset variations}
\label{tab:openness_table}
\end{table}
\section{Models}
\label{sec:models}

In this study, for the sake of brevity we focus on the high openness MLOS task, as it is the more challenging scenario. Therefore in the following we use $D_k$ to denote the set of known classes seen during training, and $D_u$ for the set of unknown classes seen only during inference.

In this section, we present five baseline models, two of which use oracle sources as a way of further exploring the limitations of these approaches. 

\subsection{Multi-label }
\label{sec:multilabel}

Given a multi-label input example $x$, the classifier $C$ generates a logit vector $\mathbf{v} = C(x) \in \mathbb{R}^{N}$, where $N := |D_k|$ i.e. KK classes present during training. To estimate whether the input contains a class in $D_k$, we take the indices above a threshold $\lambda$, i.e. $\{j : v_j > \lambda ; j \in [0, N-1] \}$.

Our baseline approach to the MLOS task is to run inference using a standard multi-label classifier. Then, to predict the unknown class we use the open-set decision criteria discussed later in this section.

The classifier consists of two stages. The first stage is a frozen OpenL3 encoder pre-trained on the environmental subset of Audioset \cite{cramer_look_2019}, which has shown competitive performance across a variety of audio and music classification tasks in the NeurIPS HEAR 2021 challenge \cite{turian2022hear}. The encoder input is a 256 frequency bin log-melspectrogram input, with output embeddings of dimension 6144. 

The second stage is a multi-layer perceptron (MLP) with five dense layers. 
Each layer consists of 1024 units and ReLU activation. The number of output units depends on the number of classes in the dataset variant, i.e. $\vert D_k\vert$ classes.  This system is depicted in Figure \ref{fig:multilabel-model}.

The multi-label classifier output has sigmoid activations and is trained using binary cross-entropy loss. Instead of using a threshold in our experiments, we used an overly-optimistic oracle strategy, picking the $m$ sources with the highest logits, where $m$ is the polyphony from the ground-truth data. We use the checkpoint with the best validation loss for evaluation.

\subsection{Combinatorial multi-class}
\label{sec:combinatorial}

In order to isolate the effect of multi-label training, we include a `combinatorial multi-class' model. Here we map each unique label combination in the OST training set to a class ID, effectively creating a multi-class model training setup. While OST has around 8000 unique class combinations, we note that this approach would lead to a `combinatorial explosion' and may be infeasible as the number of classes and unique combinations increase. 

Apart from a categorical cross-entropy loss function and different number of output layer units, we use the same architecture and training setup as described in Section \ref{sec:multilabel}.

\subsection{Source estimates multi-class PIT}
\label{sec:estimates-pit}

Since prior work on open-set AC has been in the multi-class setup, we include a model with a universal source separation front-end module to convert the MLOS task to a set of multi-class open-set classification tasks and leverage existing approaches for these subtasks. Related prior work successfully used such a universal source separation pre-processing step to improve classification precision in multi-label closed-set birdsong classification \cite{denton2022improving}. 
A separation model generates source estimates for a multi-class classifier that generates predictions. We hypothesize that the separation model will also improve performance on the MLOS task, particularly on unseen known class combinations (which could be misclassified by an open-set model as \textit{unknown} if estimated as a whole) and for clips with high polyphony. However, this approach does come with the risk of error propagation from the separation model to the classifier caused by poor source estimates. 

Given an input example $x$, the separation model $S$ generates eight source estimates $s_i$. 
Using $m$ of these eight source estimates as input, the multi-class classifier $C$ generates logits $\mathbf{v}_{i} = C(s_i) \in \mathbb{R}^N$ for each source $s_i$, where again $N := |D_k|$ , and we use the class with the max logit as our known class prediction for that source, i.e. $\operatorname{argmax}(\mathbf{v}_i)$.

\begin{figure}
    \centering
    \includegraphics[width=\columnwidth]{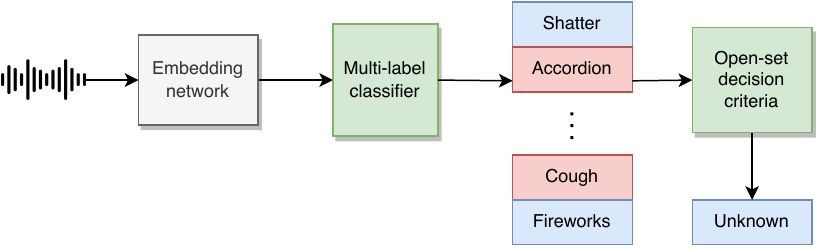}
    \caption{Multi-label model consisting of a pre-trained frozen OpenL3 embedding network and a MLP classifier.}
    \label{fig:multilabel-model}
\end{figure}
\begin{figure}
    \centering
    \includegraphics[width=\columnwidth]{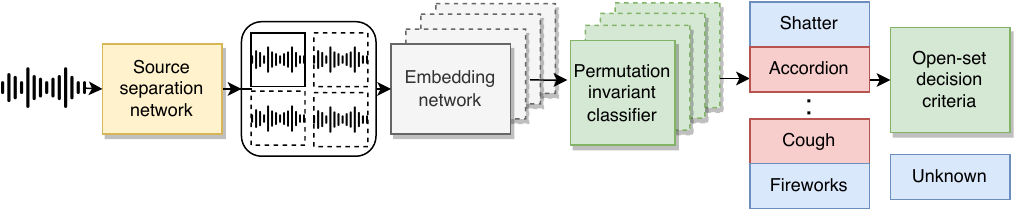}
    \caption{Source estimates multi-class PIT model consisting of a separation network and multi-class classifier trained using a permutation invariant loss. The separation network is trained separately using MixIT, then its weights are frozen as the classifier is trained. }
    \label{fig:estimates-pit-model}
\end{figure}

The separation network is a TDCN++ model trained on unlabeled polyphonic mixtures using mixture invariant training (MixIT) \cite{wisdom2020unsupervised}. As the authors of \cite{denton2022improving,wisdom2020unsupervised} note the importance of training MixIT on the target domain for quality source estimates, we train from scratch on data from all variants of the OST dataset for 1M steps and use the checkpoint with the best validation performance. 
Estimating the number of actual sources from the 8 fixed outputs is a challenging task and a potential failure point. In this paper, we opt for an overly optimistic scenario and use an oracle pruning strategy for testing. We pick the $m$ source estimates with the highest energy, where $m$ is the number of ground-truth sources. We follow this protocol both during training and inference. An existing risk of this approach is that the chosen source estimate may only contain background for input examples with low SNR. Additionally, this protocol may be sub-optimal if the model over-separates, especially in examples with low polyphony. 

The multi-class classifier has the same architecture as in Section \ref{sec:multilabel} , and is trained using a permutation invariant cross-entropy loss \cite{wisdom2020unsupervised}. Since the label assignment is only available at the clip level, we generate a prediction for each source estimate and compute the total loss for $m!$ label-source combinations. The best match that minimizes the total loss is used to update the model weights. We use the suffix permutation invariant training (PIT) to denote that a model is trained this way. The model is depicted in Figure \ref{fig:estimates-pit-model}.

\subsection{Oracle sources multi-class PIT}
\label{sec:oracle-pit}
In order to understand the effects of error propagation due to the separation network, we train a model with a perfect separation model, i.e. with the oracle sources. These oracle sources when re-combined yield the OST clips used to train the multi-label classifier model. We use the same model and training setup as in Section \ref{sec:estimates-pit}.

\subsection{Oracle sources multi-class model}

A key limitation of PIT is that it does not guarantee accurate source-label matching during training. In order to further isolate the effect that PIT may have on performance, we evaluate a reference multi-class model with the same architecture trained with oracle sources using standard cross-entropy loss. Given our modeling choices, this serves as an expected upper bound in terms of performance, as it is a true multi-class model.

\subsection{Open-set decision criteria}

We evaluate two simple open-set decision criteria that have been used previously in multi-class open-set studies. Here, we use these techniques both in the multi-class and multi-label configurations, however, the latter would suffer from false positives in scenarios with no activity or background noise events.

The first approach is softmax thresholding, where the maximum softmax probability (MSP) is compared against a threshold $\delta$ \cite{dubuisson1993statistical}-- where a model predicts \textit{unknown} if it is below and \textit{known} otherwise. Let $\hat{\mathbf{y}}$ be the classifier output for models without separation, e.g. $\hat{\mathbf{y}}= \operatorname{sigmoid}(\mathbf{v})$, and $\hat{y}_o \in \{0, 1\}$ the open-set prediction, with 0 and 1 denoting a known and unknown class prediction respectively, then 
\begin{equation}
\label{eqn:no-sep-msp}
\hat{y}_o = \begin{cases} 1 \text{ if } \max(\hat{\mathbf{y}}) < \delta; \text{ else } 0\end{cases}
\end{equation}
For PIT models and the oracle sources multi-class model, we predict unknown if any of the \textit{m} source estimates contain an unknown class:
\begin{equation}
\hat{y}_o = \begin{cases} 1 \text{ if } \max(\hat{\mathbf{y}}_i) < \delta, \text{ for } i \in [0, m-1]; \text{ else } 0
\end{cases}
\end{equation} where $\hat{\mathbf{y}}_i$ is the classifier output for a source estimate.

The second approach is Openmax \cite{bendale_towards_2016}, which aims to correct `overconfident' model predictions when the example is less likely to belong to the training distribution of the predicted class. Openmax re-weights the logit vector by penalizing the top $\alpha$ ranked logits using models of the training distribution tail for each class. The class-specific models are parameterized by the Weibull distribution tail size $\tau$ and logit rank limit $\alpha$. It also computes an unknown class probability $p_u$ based on the degree of recalibration needed, which is then appended to the updated classifier output. We refer the reader to \cite{bendale_towards_2016} for further details. 

For models without separation, we compute the updated classifier output $\hat{\mathbf{y}}_w$ using the re-weighted logit vector $\mathbf{v}_w$, e.g. $\hat{\mathbf{y}}_w= \operatorname{sigmoid}(\mathbf{v}_w)$. Then, similar to Equation \ref{eqn:no-sep-msp}--
\begin{equation}
\hat{y}_o = \begin{cases} 1 \text{ if } \max(\hat{\mathbf{y}}_w) < \delta \text{ or} \max(\hat{\mathbf{y}}_w)=p_u; \text{ else } 0
\end{cases}
\end{equation} For models with separation we apply this re-weighting and thresholding protocol to $\mathbf{v}_i$, the source estimate logit vectors.

We tune $\delta$, $\tau$, and $\alpha$ on the tuning validation set using Optuna, a Python package for efficient hyperparameter optimization \cite{optuna_2019}, and use hyperparameters from the trial that maximizes unknown detection accuracy.







\section{Evaluation}
\label{sec:results}
\begin{table}[]
\centering
\begin{tabular}{lcll}
& \multicolumn{2}{c}{\textbf{Accuracy} (SD)}                      \\ \cline{2-3}
\multicolumn{1}{c}{\textbf{}} & MSP  & Openmax         \\ \hline
Multi-label                    & 57.4 (2.9)    &          --      \\
Source estimates PIT          & 54.3 (3.0) &       --       \\
Oracle sources PIT            & \textbf{59.7} (4.7)   & \textbf{61.3} (3.0) \\
Combinatorial multi-class      & 59.1 (1.5) &       --       \\
Oracle sources multi-class     & \textbf{61.1} (3.8)    & \textbf{61.2} (3.8) \\ \hline
\end{tabular}%
\caption{Unknown detection results using maximum softmax probability thresholding (MSP) and openmax. All results are accuracy averaged over the five dataset variants, with standard deviations in parentheses.}
\label{tab:unk-results}
\end{table}


We evaluate the models separately on closed-set classification and unknown detection. For the former, we evaluate the model only on examples without unknown classes. For the latter, we evaluate the models on all examples at the clip level for a binary classification task. We present the unknown detection results in Table \ref{tab:unk-results} and closed set classification results in Table \ref{tab:kk-results}. 

From Table \ref{tab:unk-results}, we note that the multi-label model is worse than the oracle sources multi-class model. 
In this dataset, every example has at least one source, however, in scenarios where no event may be present we expect this gap to be larger, as the multi-label model may generate more false positives during silence or background noise events.

Combinatorial multi-class is only slightly worse than oracle sources multi-class. While this is an interesting finding, there are two key limitations. This model does not scale well as the number of classes increases, leading to the `combinatorial explosion' issue \cite{phan2022polyphonic}. Furthermore, this dataset follows the imbalanced source dataset distribution making certain known classes more likely than others, meaning that the model does not encounter new class combinations in the test set, leading to an optimistic view of its unknown detection accuracy. We expect this model to perform poorly in scenarios with unseen combinations of known classes, potentially generating false positives.

Oracle sources PIT does better than the multi-label model by about 4\%, which suggests that a perfect universal source separator could improve performance on this task. However, the gap is smaller than expected, potentially due to false positives caused by overconfident model predictions \cite{bendale_towards_2016}. We see some evidence of this in Table \ref{tab:unk-results} where Openmax accuracy for the oracle sources PIT model is better than its MSP accuracy, suggesting that this model is falsely overconfident for examples containing unknown class events. 

We also note that oracle sources multi-class is better than oracle sources PIT by about 2\%-- since they are both trained on the same data, the difference must be due to PIT.

%
Finally, source estimates PIT is not as good as the oracle sources PIT model, and in fact, performs worse than the multi-label model. This indicates that more research may be needed for universal source separation models to be useful in this task. Some prior results suggest that training the classifier together on the input mixture and source estimates may improve closed-set classification \cite{denton2022improving}, but it remains to be seen whether this translates to unknown detection where the model needs to separate out unknown class events as well.

\begin{table}[]
\centering
\resizebox{\columnwidth}{!}{%
\begin{tabular}{lllcll}
\multicolumn{1}{c}{\textbf{}} & \textbf{Micro F1} & \textbf{Macro F1} &  \textbf{mAP} \\ \hline
Multi-label                & 0.449 (0.01) & 0.349 (0.02) & 0.400 (0.02) \\ 
Source Estimates PIT      & 0.407 (0.01) & 0.332 (0.01) & 0.347 (0.01) \\
Oracle Sources PIT        & \textbf{0.511} (0.02) & \textbf{0.461} (0.04) & \textbf{0.501} (0.04) \\
Oracle sources multi-class & \textbf{0.581} (0.01) & \textbf{0.541} (0.01) &  \textbf{0.590} (0.01)            \\ \hline
\end{tabular}%
}

\caption{Closed-set classification results on 53 or 54 classes, depending on the dataset variant. All metrics are averaged over the five dataset variants, with standard deviations in parentheses.}
\label{tab:kk-results}
\end{table}


We notice similar trends in closed-set classification (Table \ref{tab:kk-results}) as in unknown detection MSP accuracy. 
The multi-label model as well as the oracle sources PIT model perform significantly worse than the oracle sources multi-class model, which is in line with the expectation of multi-label classification being a more challenging task.
Oracle sources PIT does better than the multi-label model, which suggests that a perfect source separation model would be useful. Lastly, the overall modest performance of the oracle sources multi-class model on both closed- and open-set tasks suggests that better audio representations are also needed to improve performance. 

\section{Discussion and Conclusion}
\label{sec:discussion}
In this work, we introduced the multi-label open-set audio classification (MLOS) task and developed a synthetic dataset with varying unknown class distributions. We then presented several baseline models using combinations of existing machine listening techniques and evaluated their performance on known class and unknown class metrics. 

We show that MLOS is a challenging task that existing approaches alone cannot adequately solve. In our study, we find that a perfect source separation model may be useful for MLOS, but further research is needed for universal source separation models to provide similar improvements in open-set classification. 

While we see some interesting results, some other questions were raised, such as how unseen known class combinations might affect unknown class detection, particularly for the multi-label and combinatorial multi-class models. We plan to evaluate this by varying vocabulary and dataset size to control the ratio of seen and unseen known class combinations in the test set.


Moreover, we consider here a simplistic data scenario where there is always at least one sound present. We plan to investigate how the inclusion of background event classes would affect some of the models discussed here, such as the multi-label and source estimates multi-class PIT.

By sharing the dataset and these baseline results, we hope to invite further interest from the community to this under-explored area of research.




\section{ACKNOWLEDGMENT}
\label{sec:ack}
We would like to thank Simran Kaur for helping to generate the initial version of the dataset.

\newpage
\bibliographystyle{IEEEtran}
\bibliography{refs}

%
%
%
%
%
%
%
%
%

\end{sloppy}
\end{document}